\documentclass{amsart}

\usepackage{amsthm}
\usepackage{amssymb}
\usepackage{amsmath}
\usepackage{graphicx}

\newtheorem {Theorem}   {Theorem}

\numberwithin{Theorem}{section}

\newtheorem {Lemma}[Theorem]    {Lemma}

\theoremstyle{definition}

\theoremstyle{remark}

\expandafter\chardef\csname pre amssym.def at\endcsname=\the\catcode`\@
\catcode`\@=11
\def\undefine#1{\let#1\undefined}
\def\newsymbol#1#2#3#4#5{\let\next@\relax
 \ifnum#2=\@ne\let\next@\msafam@\else
 \ifnum#2=\tw@\let\next@\msbfam@\fi\fi
 \mathchardef#1="#3\next@#4#5}
\def\mathhexbox@#1#2#3{\relax
 \ifmmode\mathpalette{}{\m@th\mathchar"#1#2#3}%
 \else\leavevmode\hbox{$\m@th\mathchar"#1#2#3$}\fi}
\def\hexnumber@#1{\ifcase#1 0\or 1\or 2\or 3\or 4\or 5\or 6\or 7\or 8\or
 9\or A\or B\or C\or D\or E\or F\fi}

\font\teneufm=eufm10
\font\seveneufm=eufm7
\font\fiveeufm=eufm5
\newfam\eufmfam
\textfont\eufmfam=\teneufm
\scriptfont\eufmfam=\seveneufm
\scriptscriptfont\eufmfam=\fiveeufm

\catcode`\@=\csname pre amssym.def at\endcsname

\def    \eps    {\epsilon}

\newcommand{\CH}{{\mathcal H}}

\newcommand{\CO}{{\mathcal O}}

\def    \R      {{\mathbb R}}

\def    \p      {\partial}

\begin{document}





\title[Collision Orbits]{Collision Orbits for a Hill's Type Problem}

\author[C\'esar Castilho]{C\'esar Castilho}
\address{Departamento de Matem\'atica, Universidade Federal de Pernambuco,
Recife, PE, CEP 50740-540, Brazil}

\email{castilho@dmat.ufpe.br}

\date{July, 2001}

\thanks{This work was supported by  CNPq-Brazil}

\bigskip

\begin{abstract}
 We study the planar problem of two satellites attracted by a center of
force. Assuming that the center of mass of the two-satellite system is
on a circular orbit around the center of force and using Levi-Civita
regularization we prove the existence of an almost periodic orbit
with an infinite number of collision between the satellites.
\end{abstract}
\maketitle
\centerline{ {\bf Key Words:} Hill's Problem, Regularization, Collisions.}
\section{Introduction}
    Consider the planar problem of three bodies of masses $m_0$,$m_1$ and
$m_2$ in the case where $m_1$ and $m_2$ are much smaller than $m_0$. The
mutual atraction
of the two small bodies can be usually neglected and the problem reduces
in
a fair approximation to two independent two-body problems. However
if the distance between the two small bodies is small their mutual
atraction can no longer be ignorated. This is known as Hill's problem
\cite{hill}.
The derivation of Hill's equations usually found in the literature
assumes a hierarchy of masses for the three bodies:
$$ m_0 >> m_1 >> m_2, $$
and proceeds in two steps: First the limit $m_2 \rightarrow 0$ is taken,
which gives the restricted three-body problem; then the limit $m_1
\rightarrow 0$ is taken. \par
 A number of problems in celestial mechanics can be approximated by Hill's
equations. Examples are: the Sun-Earth-Moon problem and the interaction
between satellites on nearby orbits. The nearby orbits problem was our
motivation for this work. From the data obtained during
the passage of {\it Voyager} near Saturn, it was
found that the two satellites {\it Epimatheus} and {\it Janus} around
Saturn have almost circular orbits, with radii 151,422 Km and 151,472
Km, and periods 16.664 and 16.672. {\it At each close
encounter, they interchange orbits}. To explain this behaviour some models
have been proposed, among them Scaling Techniques \cite{cors} and
Assymptotic Expansions \cite
{henon}. A common feature of the proposed
models is an assymptotic expansion in the variable that represents
the distance between the two satellites and truncation of higher order
terms. In other words, such models assume the hyphothesis that
there is a smaller bound for the distance among the satellites.
In this paper we show that this hypotesis can not be assumed without
further assumptions. Using Levi-Civita
regularization we show the
existence of an almost periodic orbit with an infinite number of
collisions among the satellites. Therefore, it's not true in general
that such lower bound exist.\par
 We take a different approach to Hill's problem. First, instead
of taking the limits $m_1, m_2 \rightarrow 0$ we fix the body
of mass $m_0$ at the
origin and assume $m_0=1$. We take $m_1,m_2 << 1$ but
no hierarchy for the masses $m_1$ and $m_2$ will be assumed.
Second, we assume that the center of mass of the two-satellite
system is on a circular orbit around the center of force. This
is equivalent to consider the circular Hill problem
\cite{itch}. The collision
between the satellites is them regularized using the canonical
form of the Levi-Civita regularization \cite{SS}. The organization
of the paper is as follows:
\par
 In section 2 we develop our model for the Hill's Problem. The model is
regularized and contrary to the usual study of Hill's problem we
do not use a rotating system of coordinates (synodical). We show that the
potential is symmetric with respect to
reflections about the origin in the regularized physical plane. This
will be crucial to prove existence of the collision orbit.
\par In section 3 we study the collinear equilibria predicted
by our model. It's an interesting feature of the model that
the regularized equations allows to find the
 equilibria points as the roots of a polynomial
of degree 4. These equations are solved and the critical
colinear points are found to be unstable. \par
 In section 4 we use a continuity argument to prove the existence
of an almost periodic orbit with an infinite number of collisions. The
basic idea is to use the reciprocal of the distance of the center
of mass of the two-satellite system to the center of force as a
perturbation parameter $\eps$. For $\eps=0$, the
binary made by the satellites will be at an infinite distance of the
center of force and the problem is integrable. Since
equations are regularized, the Kepler equations for the binary are now
represented by a ressonant harmonic oscillator (RHO). For small $\eps$,
solutions of the perturbed problem are close to the solutions of the
RHO. But solutions of the RHO that pass trough the origin are
segments of straigth lines. Perturbing these solutions we expect
that at least one of the perturbed solutions will preserve the
feature of passing through the origin more then one time. This is the
content of our main lemma (\ref{main}). The symmetry of the Hamiltonian
with respect to
the origin guarantees the existence of the collision orbit.  \par
\subsection*{Acknowledgments.} I am grateful to Jair Koiller for
introducing me to the two-satellite problem. It's also a great pleasure to
thank Hildeberto Cabral, Claudio Vidal and Marcelo Marchesin for
many discussions and suggestions. I would like to thank CNPq-Brazil for
financial support.

\section{The Model}
The Hamiltonian $\bar H$ of the planar problem of two bodies of mass $m_1$
and $m_2$
attracted by a center
of force of mass $m_0$ at the origin is
$$ \bar H = \frac{{\bar p_1}^2}{2 m_1} + \frac{{\bar p_2}^2}{2 m_2} -
G\frac{m_0
m_1}{|\bar q_1|}
- G\frac{m_0 m_2}{|\bar q_2|} - G\frac{m_1 m_2}{|\bar q_1 -  \bar q_2|} $$
where $\bar q_1$ and $\bar q_2$ are the coordinates of the bodies of
masses
$m_1$ and $m_2$ respectively, $\bar p_1$ and $\bar p_2$
their conjugate momenta and $G$ is the gravitational constant. We choose
units such that $G=1$ and set $m_0=1$. This Hamiltonian
represents a Hill problem when $ m_1,m_2 << 1$. We introduce a proportionality
factor $\lambda \in \left( 0, \infty \right)$ such that $m_2=\lambda m_1$.
 The Hamiltonian becomes
$$ \bar H = \frac{{\bar p_1}^2}{2 m_1}  + \frac{{\bar p_2}^2}{2 \lambda
m_1} -
\frac{m_1}{|\bar q_1|}
-\frac{\lambda m_1}{|\bar q_2|} -\frac{\lambda \,m_1^2}{|\bar q_1 - \bar
q_2|}. $$
Let $\bar w = d \bar{q_1} \wedge d\bar p_1 + d \bar{q_2} \wedge d\bar
p_2 $ denote the standard symplectic 2-form. Let $X_{\bar H}$ be the
Hamiltonian vector field generated by $\bar H$.
Consider the fiber scaling given by
$$\Phi(\bar q_1, \bar q_2, \bar p_1, \bar p_2)=(q_1,q_2,m_1 p_1, m_1
p_2).$$
Under this scaling we have
$$\bar H = m_1\,\left\{ \frac{p_1^2}{2} + \frac{p_2^2}{2\lambda}
-\frac{1}{|q_1|} -\frac{\lambda}{|q_2|} - \frac{\lambda m_1}{|q_1 - q_2|}
\right\};$$
and
$$ \bar w = m_1\left( dq_1 \wedge dp_1 + dq_2 \wedge dp_2 \right).$$
Dividing Hamilton's equations $i_{X_{\bar H}}{\bar w} =d \bar H$ by $m_1$
we
see that it suffices to study the Hamiltonian flow
given by  the Hamiltonian
\begin{equation}
\label{hamil}
 H = \frac{p_1^2}{2 }  + \frac{p_2^2}{2 \lambda } -
\frac{1}{|q_1|}-\frac{\lambda}{|q_2|} -\frac{\lambda \, m_1}{|q_1 - q_2|}
\end{equation}
with standard symplectic 2-form $w = dq_1 \wedge dp_1 + dq_2 \wedge
dp_2.$ \par
We introduce Jacobi variables $\rho$ and $r$ by
$$ \left\{ \begin{array}{l}
             q_1=\rho - \frac{\lambda}{1 + \lambda} \, r ,\\
              \phantom{.} \\
             q_2=\rho + \frac{1}{1 + \lambda} \, r.  \end{array} \right.
$$
Here $\rho$ represents the position of the center of mass of the
two satellites and $r$ represents their relative position vector.
The Hamiltonian (\ref{hamil}) becomes
\begin{equation}
\label{hamil1}
 H = \frac{p_{\rho}^2}{2\,(1+\lambda)} +
\frac{p_r^2}{2\,\Gamma}
-\frac{1}{|\rho -\frac{\lambda}{1 + \lambda} \, r |}
-\frac{\lambda}{|\rho + \frac{1}{1 + \lambda}\, r|} -\frac{\lambda \,
m_1}{|r|}
\end{equation}
where $p_{\rho}$ and $p_r$ are the momenta canonicaly conjugate
to $\rho$ and $r$ respectively and $\Gamma = \frac{\lambda}{1+\lambda}$.
Assuming $\frac{|r|}{|\rho|}
< \frac{1 + \lambda}{\lambda} $
we have the convergent expansions \cite{Br}
\begin{equation}
\label{expand}
   \frac{1}{|\rho \pm \frac{\lambda}{1 + \lambda}\,r |} =\frac{1}{|\rho|}
\sum_{n =
0}^{\infty} P_n(\cos \theta )\left(\mp \frac{\lambda}{1+\lambda} {\frac{|
r|}{|\rho|}}\right)^n,
\end{equation}
where $P_n(x)$ is the n-th Legendre polynomial and $\theta $ is the
positively oriented angle between $\rho$ e $r$. Hamiltonian (\ref{hamil1})
can be written as
\begin{equation}
\label{hamil2}
   H =\left( \frac{p_{\rho}^2}{2\,\bar{\lambda}} -
\frac{\bar{\lambda}}{|\rho|}\right) +
\left( \frac{p_r^2}{2\,\Gamma } -
\frac{\lambda \, m_1}{|r|}\right) -
\frac{1}{|\rho|} \sum_{n=1}^{\infty} P_{n}(\cos \theta
)\left({\frac{|r|}{|\rho|}}\right)^{n}\, \Lambda_n.
\end{equation}
where
$$ \Lambda_n = \Gamma^n \left( 1 + (-1)^n \, \lambda \right) ,$$
and $\bar{\lambda} = 1 + \lambda$.

The first parenthesis term of (\ref{hamil2}) represents the Kepler problem
described by
the center of mass around the center of force and the second parenthesis
term of (\ref{hamil2}) represents the Kepler problem of the
two satellites around their center of mass. At this point we make the
principal assumption of this work, namely, we assume that
$\rho = \left( \rho_x,\rho_y \right)$, the
vector representing the position of the center of mass of the two
satellites, describes a circular keplerian orbit of radius $|\rho_0|$
around the center of force, i.e. $\rho$ is a circular solution of
$ \ddot \rho = -\frac{1}{|\rho|^3} \rho $ yielding
$$ \rho = |\rho_0| \left( \cos (\omega t), \sin (\omega t) \right),$$
where $ \omega = |\rho_0|^{-\frac{3}{2}}$. This is equivalent to
consider the Circular Hill's Problem. By the second law of Kepler
the energy of the center of mass is given by
$E_{cm}=-\frac{\bar{\lambda}}{|\rho_0|}$, and Hamiltonian
(\ref{hamil2})
becomes
\begin{equation}
\label{hamil3}
 H = -\frac{\bar{\lambda}}{|\rho_0|} +
\left( \frac{p_r^2}{2 \Gamma} - \frac{\lambda m_1}{|r|}\right) -
\frac{1}{|\rho_0|} \sum_{n=1}^{\infty} P_{n}(\cos \theta
)\left({\frac{|r|}{|\rho_0|}}\right)^{n}\, \Lambda_n.
\end{equation}
We remark that this Hamiltonian is time dependent since the angle $\theta$
depends explicitely on time. For future reference we write
\begin{equation}
\label{tempo}
\cos(\theta )= \frac{ r_x \, \rho_{ x} + r_y \, \rho_{ y}}{|r| \, |\rho
|} = \frac{ r_x \cos (\omega t) + r_y \sin (\omega t)}{ |r|}.
\end{equation}
Since energy of system (\ref{hamil3}) is not preserved we extend phase
space from $\R^4$
to $\R^6$
by including the canonically conjugated pair $(E,t)$. Our new Hamiltonian
system is given by
\begin{equation}
\label{sist}
           \left\{ \begin{array}{l}
          \bar{\CH} = -E  -\frac{\bar{\lambda}}{|\rho_0|} +
\left( \frac{p_r^2}{2\, \Gamma} - \frac{\lambda m_1}{|r|}\right) -
\frac{\bar{\lambda}}{|\rho_0|} \sum_{n=1}^{\infty} P_{n}(\cos \theta
)\left({\frac{|r|}{|\rho_0|}}\right)^{n} \, \Lambda_n, \\
          w= du \wedge dp_u + dv \wedge dp_v + dE \wedge dt,  \end{array}
\right.
\end{equation}
where we must restrict our attention to the level set $\bar{\CH}=0.$
Denoting the new time by $f$ it follows from Hamilton's equation
$\frac{dt}{df} = 1$. By choice we identify
$f$ and $t$.
\section{Regularization}
\label{secao3}
 We regularize the collision between the two satellites. Writing
$\rho = \left( \rho_x , \rho_y \right)$ and $r = \left( r_x , r_y \right)$
we write the Levi-Civita transformation \cite{SS}
\begin{equation}
\label{levi}
           \left\{ \begin{array}{l}
             r_x = u^2 - v^2, \\
             r_y = 2 \, u \, v, \\
             \rho_x= w^2 -z^2, \\
             \rho_y= 2\, w \, z.   \end{array} \right.
\end{equation}
Identifying $r$ and $\rho$ with the complex vectors $r_x + i\, r_y$ and
$\rho_x + i \, \rho_y$ respectively, we have that  transformation
(\ref{levi}) can be written as $r = \left( u +  i \, v \right)^2$ and
$ \rho = \left( w + i \, z \right)^2$. This transformation halves
angles and therefore takes the angle $\theta$ between $\rho$ and $r$ to
its half. Writing
$\xi =(u,v)$ and $\gamma =(w,z)$ the transformation takes
$|r|$ and $|\rho|$ to $|\xi|^2$ and $|\gamma|^2$ respectively.
Observe that under this transformation the curve
$\rho(t)=|\rho_0|\left( \cos(wt), \sin(wt) \right)$ becomes
$\gamma(t)=|\rho_0|^{\frac{1}{2}}\left( \cos(wt/2), \sin(wt/2) \right)$.
 Considering the lift of (\ref{levi})
to the cotangent bundle, Hamiltonian (\ref{sist}) becomes
\begin{equation}
\label{hamil5}\bar{\CH} =-E -\frac{\bar{\lambda}}{|\gamma_0|^2} +
\frac{1}{|\xi|^2}(\frac{p_{\xi}^2}{8\,\Gamma}
-\lambda \, m_1)-
\frac{1}{|\gamma_0|^2} \sum_{n=1}^{\infty} P_{n}(\cos (\theta
/ 2)
)\left({\frac{|\xi|}{|\gamma_0|}}\right)^{2n}\, \Lambda_n . \end{equation}

Observe that $ w = |\rho_0|^{-\frac{3}{2}}=|\gamma_0|^{-3}$
Regularization is achieved  doing the time
reparametrization given
by
$$\frac{dt}{ds} = |\xi|^2 $$
where $s$ denotes the new independent variable. This reparametrization
can be performed considering the Hamiltonian
\begin{equation}
\label{hamil6} \CH= |\xi|^2 \bar{\CH}
\end{equation}
in extended phase space with symplectic 2-form given by
\begin{equation}
\label{symp}
 w = du \wedge dp_u + dv \wedge dp_v + dt \wedge dE .
\end{equation}
Since the hypersurfaces $\left\{ \bar {\CH}=0 \right\}$ and  $\left\{
\CH =0 \right\}$ are equal it follows that
the Hamiltonian flow of (\ref{hamil5}) at the level set $\bar{\CH}=0$ is
a
reparametrization of the Hamiltonian flow of (\ref{hamil6}) at the level
set
$\CH =0$. (\ref{hamil5}) and (\ref{hamil6}) yields
$$ \CH = - \lambda \, m_1 + \frac{p_{\xi}^2}{8\, \Gamma} -
\left(\frac{\bar{\lambda}}{|\gamma_0|^2} + E \right) |\xi|^2
-\frac{|\xi|^2}{|\gamma_0|^2}\sum_{n=1}^{\infty}
P_{n}\left(\cos
(\theta / 2)\right) \left(\frac{|\xi|^{2n}}{|\gamma_0|^{2n}}\right)\,
\Lambda_n.
$$
We are interested on the flow of $\CH$ at the level $0$. We can
eliminate the constant $-\lambda \, m_1$ of the Hamiltonian by considering
the level $\lambda \, m_1$ instead.
The square of the reciprocal of the radius of the center of mass
 will be treated as a
perturbation  parameter. Writing $  \epsilon =
\frac{1}{|\gamma_0|^2} $ and doing
the symplectic scaling $p_{\xi} \rightarrow 2 \, p_{\xi}$ ,
$ \xi \rightarrow \xi/2$ we have
\begin{equation}
\label{hamilre}
 \CH_{\lambda \, m_1} =
\frac{p_{\xi}^2}{2\, \Gamma} -
\frac{1}{4} \left( \bar{\lambda}\,\epsilon +E \right) |\xi|^2
-\epsilon^2 \,
\sum_{n=1}^{\infty} \epsilon^{n-1} \,P_{n}\left(\cos (\theta / 2
)\right) \left(\frac{|\xi|^2}{4}\right)^{n+1} \, \Lambda_n,
\end{equation}
where the subscript $\lambda \, m_1$ is a reminder that we must consider
the level set $\CH = \lambda \, m_1$.\par
\begin{Lemma}
\label{sym}
Hamiltonian (\ref{hamilre})
is invariant with respect to the symmetry $S: \R^6 \rightarrow \R^6$
given by $S(\xi, p_{\xi}, E, t) = (-\xi,  p_{\xi}, E, t)$, i.e.
$\CH \left(\xi,p_{\xi},E,t \right)
= \CH \left(-\xi,p_{\xi},E,t \right).$ \end{Lemma}

\begin{proof} Since $|\xi|$ is invariant under $S$ it suffices
to show that $ \cos \left(\theta /2 \right)$
is also invariant. But from (\ref{tempo}) and (\ref{levi})
$$\cos(\theta / 2)=\frac{ \left(u^2 - v^2\right) \cos(wt/2) +2 u v
\sin(wt/2)}{|\xi|^2}$$
that is clearly invariant under S.
\end{proof}

The equations of motion of (\ref{hamilre}) can be written as
\begin{equation}
\label{ordem}
\left\{
\begin{array}{l}
  \ddot{\xi}=-\frac{\bar{\lambda} \, \eps+E}{2\,\Gamma}\, \xi +\eps^2 \,
\nabla V
; \\
   \phantom{   } \\
  \dot E = -\eps^2 \,\frac{\p V}{\p t} ;\\
  \phantom{   }  \\
  \dot t =\frac{ |\xi|^2}{4}, \\
\end{array} \right.
\end{equation}
where
$$ V= \sum_{n=1}^{\infty} \,\eps^{n-1} \, P_{n}\left(\cos
(\theta / 2
)\right) \left( \frac{|\xi|^2}{4} \right)^{n+1} \, \Lambda_n.$$
From the second equation of (\ref{ordem}) we have that $E(s)=E(0)+
\CO(\eps^{\frac{7}{2}})$, and we
write for future reference that
\begin{equation}
\label{as}
   \ddot{\xi}=-\frac{E_0}{2\,\Gamma}\, \xi -\eps \, 
\frac{\bar \lambda}{2 \Gamma} \, \xi + \CO(\eps^{\frac{7}{2}}) .
\end{equation}

\section{ Euler's Critical Points }

 Euler's critical points are relative equilibria of the Hamiltonian
 system representing colinear configurations. In what follows we compute
 the Euler's critical points for our model: In a colinear configuration
the two masses will be moving forming a
straigth  line with the center of force. Therefore we must have
  $\theta = 0$ or $\theta=\pi$ and $\dot \theta =0$. The first case
represents the case where
$m_1$ is between the center of force and $m_2$ and the second case
represents the case where $m_2$ is in between. Since $m_2 = \lambda m_1$
we need to work only with the case $\theta=0$.
 We introduce polar coordinates
$(l,\phi)$ in the plane $(u,v)$,
 where $l$ is the radius and $\phi$ is the angle. Observe that
 $\phi = \pi -\frac{wt- \theta}{2} $.
 Let $p_l$ and $p_{\phi}$ be the conjugate momenta to $l$
 and $\phi$ respectively.  Hamiltonian (\ref{hamilre}) becomes
\begin{equation}
\label{hampol}
  \CH_{\lambda \, m_1}=
\frac{p_l^2}{2\, \Gamma}+\frac{p_{\phi}^2}{2\,\Gamma \,l^2}
-\frac{1}{4}\left( \bar{\lambda}\,\epsilon +E \right)\, l^2
 -\eps^2 \,
\sum_{n=1}^{\infty} \, \eps^{n-1}\,
P_{n}\left(\cos ( \theta / 2) \right) \left( \frac{l^2}{4}\right)^{n+1}\,
\Lambda_n
,
\end{equation}
Hamilton's equations are
\begin{equation}
\label{polar}
  \left\{
   \begin{array}{l}
   \dot l = \frac{p_l}{\Gamma} , \\
   \dot{ p_l} =\frac{1}{2} (\bar{\lambda}\,\epsilon + E)\, l +
\frac{p_{\phi}^2}{\Gamma \, l^3}
   +2 \eps^2 l\,\sum_{n=1}^{\infty}\eps^{n-1}
  P_{n}\left(\cos ( \theta / 2 )\right)\, (n+1) \,
\left({\frac{l^2}{4}}\right)^{n}\,\Lambda_n ,\\
   \dot \phi = \frac{p_{\phi}}{\Gamma \,l^2}, \\
   \dot{ p_{\phi}} =\eps^2 \sum_{n=1}^{\infty}
   \eps^{n-1}\left( \frac{ l^2}{4} \right)^{n+1}
   D_x\,P_n\left( \cos(  \theta / 2 )\right)
   \sin ( \theta / 2) \, \Lambda_n , \\
   \dot E = -\frac{w}{4}\,\eps^2 \sum_{n=1}^{\infty}
   \eps^{n-1}\left( \frac{ l^2}{4} \right)^{n+1}
   D_x\, P_n\left( \cos(  \theta / 2 )\right)
   \sin ( \theta / 2)\, \Lambda_n,  \\
   \dot t = \frac{l^2}{4} ;  \end{array} \right.
\end{equation}
 where $ \theta / 2 = \pi  - \phi - \frac{w \, t}{2}$.
 We look for Euler's critical points of (\ref{polar}).
 The condition  $\dot \theta=0$ implies that
 $ \dot \phi = w \dot t / 2$. Using the second and last equations
of (\ref{polar}) it follows that
 $p_{\phi}=  \frac{ \,w \,\Gamma \, l^4}{8}$. This is the value  the
 angular momentum $p_{\phi}$ must have in order to keep
 the colinear shape of the configuration. Since we
 are looking for colinear
 critical points in phase space (and
 not in extended phase space!) we must find a point for which
 $\dot l =0$, $\dot p_l =0$, $\dot{p_{\phi}}=0$ and
 $\dot E =0$. For $\theta = 0$ i.e., for $\phi = wt/2$  it follows from
(\ref{polar}) that the last
 two equalities are satisfied. We write $E_0=E(0)$. The
 first equality will be satisfied
 setting $p_l=0$. It remains to find $l$ for
 which $\dot{p_l}=0$. For $\theta=0$ we have that
 $P_n\left( \cos(\frac{\theta}{2})\right)= P_n(1)=1.$
 Recalling that $ w=  \eps^{\frac{3}{2}}$ it follows that
 \begin{equation}
 \label{pot}
 \dot p_l = -\frac{\p V}{\p l},
 \end{equation}
 where the potential function $V$ is written as
 \begin{equation}
  V(l) =-\frac{1}{4}\left(\bar{\lambda}\, \eps + E_0 \right) \, l^2 +
\frac{\eps^{\frac{3}{2}} \, \Gamma \, l^6}{16} -\eps \, \frac{l^2}{4}
\left\{
\sum_{n=1}^{\infty}\left(\eps \,\frac{l^2}{4}
\,\Gamma\right)^n
+\lambda \sum_{n=1}^{\infty}\left(-\eps \,\frac{l^2}{4}
\,\Gamma\right)^n
\right\}
\end{equation}
Suming the series we obtain
\begin{equation}
\label{inner}
V(l) = -\frac{1}{4}\left(\bar{\lambda}\, \eps + E_0 \right) \, l^2 +
\frac{\eps^{\frac{3}{2}} \, \Gamma \, l^6}{16}
-\frac{\eps^2 \, l^4 \, \Gamma}{4}  \left\{
\frac{4\bar \lambda + \eps \,l^2 \, \Gamma \, (1-\lambda)}
{16 - \eps^2 \, l^4 \, \Gamma^2} \right\}
\end{equation}
We want to find non-zero solutions of $\frac{\p V}{\p l}=0$.
Making $u=l^2$ we only need to find non-zero solutions of
$\frac{\p V}{\p u}=0$. After some straightforwad computations
we see that the zeros of this equation is given by the zeros
of a polynomial of degree $4$ in $u$. Therefore we have explicit
 analytic solutions for the Inner Euler's Critical points. The
explicit solutions can be computed in an algebraic manipulator
software but are very messy to be of any use. The critical points
can be numerically computed using a Newton algorithm. We computed
the values for diferent values of $E_0$. For all of the critical
points computed we have that $\frac{\p^2 V}{\p u^2} < 0$ indicating
that the critical points are unstable.



\section{ Collision Orbits}

  Hamiltonian (\ref{hamilre}) is symmetric with respect to reflections
about the origin. This simmetry will be used to prove existence of an
almost periodic orbit that passes through the origin twice. Since the
origin in
the regularized plane represents a collision in the physical plane,
this orbit represents a periodic orbit with an infinite
number of collisions. The proof relies on the fact that for
$\epsilon=0$ and $E_0<0$, (\ref{hamilre}) represents a ressonant harmonic
oscillator. Therefore, the projections on the $(u,v)$ plane
of trajectories leaving the origin (ejection trajectories) are segments of
straight lines. We expect that when
$\epsilon$ is small, some trajectories of the perturbed system will
preserve the feature of passing through the origin at least two times.
The symmetry with respect to the origin will then imply that
this orbit  will cross the origin an infinite number of times.\par
 We set $m_1=\mu$ and assume that $m_1,\lambda \,m_1 << 1$ and $E_0 <0$.
We write $\mu =
\CO(\eps)$. More explicitly we will assume
 that there is a constant $\tilde \mu$  such that
$\mu(\eps)=\eps\,\tilde{\mu}$. We will prove the following theorem

\begin{Theorem} If $\epsilon$ is small enough, there is an almost periodic
orbit of (\ref{hamilre}) that passes through the origin.
\end{Theorem}
\begin{proof}
 For easy reference we write
 Hamiltonian (\ref{hamilre})
\begin{equation}
\label{equal}
H_{\eps \, \kappa}=p_{\xi}^2-\frac{1}{4}\,(\bar \lambda
\, \eps + E)\, |\xi|^2 -\eps^2\,V,
\end{equation}
where $\kappa = \lambda \, \mu$. We restrict our study to solutions that
at time $s=0$ leave
the origin.\par \bigskip
\noindent {\bf Definition} We call ejection solutions, solutions of the
Hamiltonian system with Hamiltonian (\ref{equal}) and symplectic 2-form
(\ref{symp})  that at  time
$s=0$ leave the origin, i.e., solutions with $\xi(0) =(0,0).$
\par \bigskip
\noindent{\bf Remark:} The set of ejection trajectories is
parametrized by a circle. In fact, for these trajectories
(\ref{equal}) implies that $|p_{\xi}(0)|^2 = \eps \kappa$ and we
can write
$p_u(0)=\sqrt{\eps \, \kappa } \cos(\alpha)$ and
$p_v(0)=\sqrt{\eps \, \kappa} \sin(\alpha)$
for
$\alpha \in \left[ 0 , 2 \pi) \right.$. \par \bigskip
 \noindent For a small $\epsilon$ and
$E_0 < 0\;$  solutions of (\ref{equal}) will be close to the
solutions of a ressonant harmonic oscillator with
period and amplitude given by
$$  T_0=\frac{2 \pi \sqrt{2\, \Gamma} }{\sqrt {|E_0|}}; \,
\mbox{ and } A =2\,\sqrt{\frac{\eps \, \kappa}{|E_0|}}$$
respectively. Let $\eta$ denote an initial condition for an
ejection trajectory, i.e. $\eta=(0,0,p_u,p_v,E(0),t(0))$.
Fixing  $E(0)$ and $t(0)$ we have that $\eta$ is uniquely
determined by $\alpha$. We write the set of ejection trajectories
at time $s$ and parameter $\epsilon$ as
$$ \phi^s_{\epsilon}(\alpha)=(u^s_{\epsilon}(\alpha),v^s_{\epsilon}(\alpha),
{p_u}^s_{\epsilon}(\alpha),{p_v}^s_{\epsilon}(\alpha),E^s_{\epsilon}(\alpha)
,t^s_{\epsilon}(\alpha)).$$ Then
$\phi^0_{\epsilon}(\alpha)=
\eta=(0,0,\sqrt{\eps \, \kappa}\cos(\alpha),
\sqrt{\eps \, \kappa}\sin(\alpha), E(0), t(0)).$\par
The following lemma shows that, for $\eps$ small enough, an ejection
trajectory with angle $\alpha$
pointing to the right half plane will cross the $v$ axis transversaly at
some time $\tau=\tau(\alpha)$.
\begin{Lemma}
 \label{main}
 Let $E_0<0$. Let $\delta$ be a
positive real number with $ \delta << 1$. Let
$\alpha \in I$, where $I= \left[-\frac{\pi}{2}
+\delta,\frac{\pi}{2}-\delta\right]$. If
 $\epsilon$ is small enough, there
exist times $\tau=\tau(\epsilon,\alpha)$ such that
\begin{equation}
\label{per}
 \frac{1}{4}\,T_0 < \tau < \frac{3}{4}\,T_0 ,
 \end{equation}
and $ u^{\tau}_{\epsilon} (\alpha)=0 $ .
Moreover $v^{\tau}_{\epsilon} (\alpha)$ is a continuous function of
$\alpha $.
\end{Lemma}

\begin{proof} We first prove existence of $\tau$.
Let $\displaystyle m =\min_{\alpha \in I} \left\{ |\cos(\alpha)|
\right\}
=|\cos(\frac{\pi}{2}-\delta)|$. We write
\begin{equation}
\label{dec}
 \phi^s_{\eps} = \phi^s_{osc} + \eps \, \psi^s
\end{equation}
where $\phi^s_{osc}$ is the harmonic oscillator flow
given by
\begin{eqnarray*}
\label{harm}
 \phi^s_{osc}(\alpha) & = & (A\,\cos(\alpha) \sin(\omega s), A
\,\sin(\alpha) \sin(\omega s),-A \omega \cos(\alpha)\,\cos(
\omega s), \\
                      &   & \mbox{                 }
-A \omega \sin(\alpha)\,\cos( \omega s),E(0),t(0)),
\end{eqnarray*}
where $\omega = \sqrt{\frac{|E_0|}{2 \, \Gamma}}$.
From (\ref{as}) $\psi^s_{\xi}$ satisfies
$$ \ddot {\psi^s_{\xi}}=-\frac{E_0}{2} \, \psi^s_{\xi}
- \frac{\phi^s_{\xi}}{2} + \CO(\eps).$$
Let $\underbar t =\frac{1}{4}\, T_0$ and $\bar t =\frac{3}{4}\, T_0$.
Then
\begin{equation}
\label{est}
 u^{\underbar t}_{\eps}(\alpha)=\sqrt{\eps \, \kappa} \, \cos(\alpha)
+ \eps \,\psi_u^{\underbar t}(\alpha);
\end{equation}
where $\psi_u^s$ is the $u$ component of $
\psi^s$. Let $J=[0,\bar \eps]$ where $\bar \eps$ is a
positive small number.
 Let  $\displaystyle M =\max_{\eps \in J}\left( \max_{\alpha
 \in I}|\psi_u^{\underbar t}(\alpha)|\right)$. We can find $\eps$ small
enough such that
\begin{equation}
\label{epss}
 \eps^{\frac{1}{2}}\,{\kappa}^{\frac{1}{2}} \, m >
\eps \,M,
\end{equation}
this implies that
$$ | \sqrt{\eps \, \kappa} \, \cos(\alpha)| > \eps |\psi_u^{\underbar t}
(\alpha)|. $$
Therefore for
 $\epsilon$ satisfying (\ref{epss}), (\ref{est}) implies
that $u^{\bar
t}_{\eps}$ has the
same sign as $\sqrt{\eps\,\kappa}\,\cos(\alpha)$, i.e., $u^{\bar
t}_{\eps}$ is positive. Analogously we show that $u^{\bar t}_{\eps}$
is negative. Therefore by continuity of the flow,
there exists  times $\tau_{\epsilon} $,
for which $u_{\eps}^{\tau_{\epsilon}(\alpha)}=0$ and satisfying
(\ref{per}).
To prove continuity it suffices to prove that at time $s=\tau$
 the projection of the flow on the $(u,v)$ plane intersects the $v$ axis
transversaly, i.e., it suffices to show that
$\dot u^{\tau}_{\eps}(\alpha) \ne 0$. But
$$ \dot u^{\tau}_{\eps}(\alpha) = -A \omega \sin(\alpha)\,\cos(\omega
\tau)
+ \eps \,\psi^{\tau}_{u}.$$
For $\eps$ small enough it follows from (\ref{epss}), using the same
argument as in the
existence part,  that
$\dot u^{\tau}_{\eps}(\alpha) \ne 0$.
\end{proof}
For notational simplicity we will write in the next lemma
$u_{\eps}^s=u(s)$ and $v_{\eps}^s=v(s)$.
\begin{Lemma}  Let  $E_0 < 0.$ Then for $\eps$ small
enough there exists $\alpha \in I$ such that
$v(\tau_u(\alpha))=0$.
\end{Lemma}
\begin{proof} Let $\tau$ as in lemma (\ref{main}). By Taylor's
Theorem it follows that there exists $c$, $0< c < \tau$  such that
$\xi(\tau)= \xi (0) + \dot \xi (0)\, \tau + \frac{\ddot \xi (0)}{2}
\, \tau^2 + \frac{\dddot \xi (c)}{6} \tau^3 $, that we write as
$$ u(\tau) = u(0) + \dot u(0) \, \tau +\frac{ \ddot u(0)}{2} \, \tau^2
 + \frac{\dddot u(c)}{6} \, \tau^3,$$
$$ v(\tau) = v(0) + \dot v(0) \, \tau +\frac{ \ddot v(0)}{2} \, \tau^2
 + \frac{\dddot v(c)}{6} \, \tau^3.$$
For an ejection trajectory we have that $u(0)=v(0)=0$ and equations
(\ref{ordem}) imply that $\ddot u(0)=\ddot v(0) =0$. Therefore
$$ u(\tau) = \sqrt{\eps \,\kappa} \cos(\alpha) \, \tau +
\frac{\dddot u (c)}{6}
\, \tau^3,$$
$$ v(\tau) = \sqrt{\eps \,\kappa} \sin(\alpha) \, \tau +
\frac{\dddot v (c)}{6}
\, \tau^3,$$
with $0 < c < \tau$. $\tau$ satisfies (\ref{per}) and by definition
$u(\tau(\alpha))=0.$
Solving the first equation for $\tau$ and substituting on
the second equation we obtain
\begin{equation}
\label{zero}
 v(\tau_u(\alpha))=\frac{\sqrt{ \eps\,\kappa}}{\dddot u(c)}
\left(\sin(\alpha) \dddot u(c) - \cos(\alpha) \dddot v(c) \right)
.\end{equation}
Now we estimate $|\dddot u(c) - \dddot v(c)|$.
From the Taylor expansions we have that
$$ |\dddot u(c) - \dddot v(c)| < \frac{6}{\tau^2}\,|\sqrt{ \eps
\,\kappa }
(\cos(\alpha)-\sin(\alpha))| + \frac{6}{\tau^3}\,|u(\tau)-v(\tau)|.$$
Using (\ref{per})  we obtain
$$ |\dddot u(c) - \dddot v(c)| < \frac{6 \sqrt{ \eps
\,\kappa }}{\pi^2}\,|E_0|
+\frac{6 \, |E_0|^{\frac{3}{2}}}{\pi^2 \sqrt{2}}\,
|v(\tau)|.$$
But, from (\ref{dec}) we have that $v(\tau)=\frac{ \sqrt{\eps \,
\kappa}}{|E_0|} \sin(\alpha) \cos(w s) + {\CO}(\eps^3)$
giving that
$$ |\dddot u(c) - \dddot v(c)| < \frac{6}{\pi^2}\,
\sqrt{ \eps \,\kappa}\,|E_0|
+\frac{6 \, |E_0|^{\frac{3}{2}}}{\pi^2 \sqrt{2}}\,
\frac{\sqrt{\eps \, \kappa}}{|E_0|^{\frac{1}{2}}}
+ {\CO}(\eps^3).$$
Therefore there exists a constant $K$ such that
$$ |\dddot u(c) - \dddot v(c)| < K\, |E_0|\, \sqrt{\eps \, \tilde{\mu}}
+ {\CO}(\eps^3).$$
We can then write
$$ \dddot v(c) = \dddot u(c) + \CO(\eps^{\frac{1}{2}}),$$
and  (\ref{zero}) becomes
$$ v(\tau_u(\alpha))=\frac{\sqrt{ \eps \, \kappa}}{|E_0|}
\left( \sin(\alpha) - \cos(\alpha)
+{\CO}(\eps^{\frac{1}{2}})\right).$$
Thus for $\eps$ small enough we can find $\alpha_1, \,\alpha_2
\in I$  such
that $ v(\tau(\alpha_1)) > 0$ and  $ v(\tau(\alpha_2)) < 0$. By
the continuity part of lemma (\ref{main}) there is $\alpha \in I$,
such that $ v(\tau(\alpha))=0$.
\end{proof}
 Therefore we proved that for $\eps$ small enough there is an angle
$\alpha$ such that the ejection trajectory with this angle will
pass through the origin in some future time $\tau(\alpha)$. But by
lemma (\ref{sym}) the system is symmetric with reflections about
the origin. Therefore, the orbit will pass an infinite number of
times around the origin proving the theorem.
\end{proof}

\label{actanle}

 \end{document}